\centerline{\bf IS QUANTUM SPACETIME INFINITE DIMENSIONAL ? }
\bigskip
\centerline{Carlos Castro}
\centerline{Center for Theoretical Studies of Physical Systems}
\centerline{Clark Atlanta University}
\centerline{Atlanta, GA. 30314}
\smallskip
\centerline{January 2000}

\bigskip

\centerline{\bf Abstract}

The Stringy Uncertainty relations, and corrections thereof, were  explicitly derived recently from 
the New Relativity Principle that treats all dimensions and signatures on the same footing and which 
is based on the postulate that the Planck scale is the minimal length in Nature in 
the same vein that the speed of light was taken as the maximum velocity in Einstein's 
theory of Special Relativity. A simple numerical argument is presented which suggests 
that Quantum Spacetime may very well be  $infinite$ dimensional. 
A discussion of the repercusions of this new paradigm in Physics is given. 
A truly remarkably simple and plausible solution of the cosmological constant problem results from 
the New Relativity Principle : The cosmological constant is $not$ constant, in the same vein that 
Energy in Einstein's Special Relativity is observer dependent. Finally, following El Naschie, 
we argue why  the observed $D=4$ world might just be an $average$ dimension over the infinite 
possible values of the Quantum Spacetime and why the compactification mechanisms  from higher to four
dimensions in String theory may not be actually the right way to look at the world at Planck scales.

\bigskip

\centerline{\bf 1. Preface} 
\bigskip

Before starting, we wish to say that readers already familiar with [1] may skip the second  section entirely. We deem  it 
absolutely necessary to repeat 
the calculations that led us to the String Uncertainty Relations, corrections thereof, and the direct link between the Regge behaviour of string 
theory with the area quantization [1]. In order to understand the main results of this work 
in section {\bf 3 } one must follow closely section {\bf 2 } . We apologize for having repeated the results of [1]. 

The New Relativity Principle that encompasses the ideas of Noncommutative {\bf C}-spaces, [1], Polydimensional Covariance [2] 
and Scale Relativity [3] offers, in addition to the straightforward derivation of the String Uncertainty Relations, a truly 
remarkable and simple solution to the cosmological constant problem : the so called cosmological constant is {\bf not} a constant. 
This is shown following the same arguments that Einstein gave when he showed that the energy was observer dependent : it is in the eye
of the beholder. Finally,  following El Naschie, 
we argue why  the observed $D=4$ world might just be an $average$ dimension over the infinite 
possible values of the Quantum Spacetime and why the compactification mechanisms  from higher to four
dimensions in String theory may not be actually the right way to look at the world at Planck scales.      
  
\bigskip
\centerline{\bf 2. Introduction : String Uncertainty Relations from the New Relativity Principle} 
\bigskip

Recently we have proposed that a New Relativity principle may be 
operating in Nature
which could reveal important clues to find the origins of $M$ theory  
[1]. 
We were forced to introduce this new Relativity principle, where all 
dimensions and
signatures of spacetime are on the same footing, to find a fully 
covariant formulation
of the $p$-brane Quantum Mechanical Loop Wave equations. This New 
Relativity Principle,
or the principle of Polydimensional Covariance as has been called by 
Pezzaglia,
has also been crucial in the derivation of Papapetrou's equations of 
motion of a
spinning particle in curved spaces that was a long standing problem
which lasted almost 50 years  [2]. A Clifford calculus was used where 
all the
equations were written in terms of Clifford-valued multivector 
quantities;
i.e one had to abandon the use of vectors and tensors and replace them 
by
Clifford-algebra valued quantities, matrices, for example .

In this section  we will explicitly derive the String Uncertainty 
Relations,
and corrections thereof, directly from the Quantum Mechanical Wave 
equations
on Noncommutative Clifford manifolds or {\bf C}-spaces  [1]. 
There was a one-to-one correspondence between the nested hierarchy of 
point, loop, {\bf 2}-loop,
{\bf 3}-loop,......{\bf p}-loop histories encoded in terms of 
hypermatrices
and wave equations written in terms of Clifford-algebra valued 
multivector quantities.
This permits us to recast the QM wave equations associated with the 
hierarchy of nested 
{\bf p}-loop histories, embedded in a target spacetime of $D$ dimensions 
, 
where the values of $p$ range  from :  $p=0,1,2,3......D-1$, as a 
$single$ QM line
functional wave equation whose lines live in a Noncommutative Clifford 
manifold of
$2^D$ dimensions. $p=D-1$ is the the maximum value of $p$ that 
saturates the
embedding spacetime dimension.

The line functional wave equation in the Clifford manifold, {\bf 
C}-space  is :

$$\int d\Sigma~( {\delta^2 \over \delta X(\Sigma) \delta X(\Sigma) } 
+{\cal E}^2 )
\Psi [X(\Sigma)]=0. \eqno (1)$$
where $\Sigma$ is an invariant evolution parameter of $l^{D}$ dimensions 
generalizing the notion of the invariant proper time in Special 
Relativity
linked to a massive point particle line ( path ) history :

$$(d\Sigma)^2 = (d\Omega_{p+1})^2 + \Lambda^{2p}(dx^\mu dx_\mu)
+  \Lambda^{2(p-1)}(d\sigma ^{\mu\nu}  d\sigma_{\mu\nu} )
+  \Lambda^{2(p-2)}(d\sigma^{\mu\nu\rho} d \sigma _{\mu\nu\rho})
+ .......\eqno (2)$$
$\Lambda$ is the Planck scale in $D$ dimensions. 
{\bf X}$(\Sigma)$  is a Clifford-algebra valued " line " living in the 
Clifford
manifold ( {\bf C}-space)  :

$$X=\Omega_{p+1} +\Lambda^p x_\mu \gamma^\mu +\Lambda^{p-1} \sigma_{\mu\nu} \gamma^\mu 
\gamma^\nu +\Lambda^{p-2}
\sigma_{\mu\nu\rho}\gamma^\mu \gamma^\nu \gamma^\rho +......... \eqno 
(3a)$$

The multivector {\bf X} encodes in one single stroke the point history 
represented by the
ordinary $x_\mu$ coordinates and the holographic projections of the 
nested family of   
{\bf 1}-loop, {\bf 2}-loop, {\bf 3}-loop...{\bf p}-loop histories onto 
the embedding
coordinate spacetime planes given respectively by : 
$$\sigma_{\mu \nu},  \sigma_{\mu \nu\rho}......\sigma_{\mu_1 
\mu_2...\mu_{p+1}}\eqno (3b)$$
The scalar $\Omega_{p+1}$ is the invariant proper 
$p+1=D$-volume associated
with the motion of the ( maximal dimension ) {\bf p}-loop across the 
$D=p+1$-dim target
spacetime. 
There was a coincidence condition [1] that required to equate the values 
of the
center of mass coordinates $x_\mu$, for all the {\bf p }-loops, with the 
values of the
$x^\mu$ coordinates of the
point particle path history. This was due to the fact that
upon setting $\Lambda=0$ all the {\bf p}-loop histories collapse to a 
point history. 
The latter history is the baseline where one constructs the whole 
hierarchy. 
This also required a proportionality relationship :

$$\tau \sim { A\over \Lambda }\sim {V \over \Lambda^2}\sim.......
\sim {\Omega^{p+1} \over \Lambda^p}. \eqno (4)$$
$\tau,A,V....\Omega^{p+1}$ represent the invariant proper time, proper 
area, proper volume,...
proper $p+1$-dim volume swept  by the  point, loop, {\bf 2}-loop, 
{\bf 3}-loop,.....
{\bf p}-loop histories across their motion through the embedding spacetime, respectively.
${\cal E}=T $ is a quantity of dimension $(mass)^{p+1}$, the maximal 
$p$-brane tension ( $p=D-1$) .

The wave functional $\Psi$ is in general a Clifford-valued, hypercomplex 
number.
In particular it could be a complex, quaternionic or octonionic valued 
quantity.
At the moment we shall not dwell on the very subtle complications and 
battles associated
with the quaternionic/octonionic extensions of Quantum Mechanics [14] 
based on Division algebras and simply take the wave function to be a 
complex number. 
The line functional wave equation for lines living in the Clifford 
manifold ( {\bf C}-spaces)
are difficult to solve in general. To obtain the String Uncertainty 
Relations, 
and corrections thereof,  one needs to simplify them.
The most simple expression is to write the simplified wave equation in units 
$\hbar=c=1$ :

$$[- ( {\partial ^2 \over \partial  x^\mu  \partial  x_\mu }
+{\Lambda^2\over 2}  {\partial ^2 \over \partial \sigma^{\mu\nu} 
\partial  \sigma_{\mu\nu}}
+ {\Lambda^4 \over 3!} {\partial ^2 \over \partial \sigma^{\mu\nu\rho} 
\partial
\sigma_{\mu\nu\rho} }
+......) - \Lambda^{2p} {\cal E} ^2 ]~\Psi [x^\mu, \sigma^{\mu\nu}, \sigma^{\mu\nu\rho},.....  ]=0 \eqno 
(5)$$
where we have dropped the first component of the Clifford multivector
dependence, $\Omega^{p+1}$, of the wave functional $\Psi$ and we have replaced 
functional differential
equations for ordinary differential equations. 
Had one kept the first component dependence   $\Omega^{p+1}$ on $\Psi$ one would have had a cosmological constant contribution to the 
${\cal E}$ term as we will see below.  Similar types of 
equations in a
different context  with only the first two 
terms of eq-(5),
have also been written in [2].

The last equation contains the seeds of the String Uncertainty Relations 
and corrections thereof.
Plane wave type solutions to eq-(5) are :

$$\Psi =e^{i ( k_\mu x^\mu + k_{\mu\nu}  \sigma^{\mu \nu} +  k_{\mu\nu\rho}  \sigma^{\mu \nu\rho}+ .......)}. \eqno (6)$$
where $k_{\mu\nu}, k_{\mu\nu\rho}.....$ are the area-momentum, volume-momentum,..... $p+1$-volume-momentum conjugate variables to the holographic 
$\sigma^{\mu\nu}, \sigma^{\mu\nu\rho}...$ coordinates respectively. These are the components of the Clifford-algebra valued 
$multivector$ {\bf K} that admits an expansion into a family of antisymmetric tensors of arbitrary rank like the Clifford-algebra valued "line" 
{\bf X} did earlier in eq-(3a). The multivector {\bf K} is nothing but the conjugate $polymomentum$  variable to {\bf X} in {\bf C}-space. 
Inserting the plane wave solution into the simplified wave equation  
yields the generalized 
dispersion relation, after reinserting the suitable powers of $\hbar$ :

$$\hbar^2  (k^2 +{1\over 2} \Lambda^2 (k_{\mu\nu} ) (k^{\mu\nu}) 
+{1\over 3!} \Lambda^4 (k_{\mu\nu\rho}  ) (k^{\mu\nu\rho} )+
........) - { \Lambda^{2p} {\cal E}^2 \over \hbar^{2p} }       =  0 . \eqno (7)$$
this is just the generalization of the ordinary wave/particle dispersion 
relationship

$$p^2 =\hbar^2 k^2 .~~~p^2-m^2=0 . \eqno (8)$$
Had one included the  $\Omega^{p+1}$ dependence on $\Psi$; i.e an extra piece $exp~[i\Omega_{p+1}\lambda ]$, where $\lambda$ is the cosmological constant 
of dimensions $(mass)^{(p+1)} $.  
The required $- \Lambda^{2p} \partial^2 \Psi / (\partial \Omega_{p+1})^2  $ 
term of the simplified wave equation (5) would have generated an extra term of the form $ \Lambda^{2p}\lambda^2 $. After reinserting the suitable powers of $\hbar$, 
the cosmological constant term will precisely $shift$ the value of the 
$- \Lambda^{2p} {\cal E}^2/\hbar^{2p} $ piece of eq-(7) to the value : $- ({ \Lambda \over \hbar} )^{2p}({\cal E}^2 - \lambda^2)$, which precisely has an overall  dimension of 
$m^2$ as expected.  

Hence, this will be then the " vacuum " 
contribution to $maximal$  $p$-brane tension ( $p=D-1$) : ${\cal E} =T_p$ has overall units $(mass)^{p+1}$; i.e energy per $p$-dimensional volume.         
On dimensional grounds and due to the $coincidence$ condition [1] referred above one has that :

$$(k_{\mu\nu}    ) (k^{\mu\nu}  ) = \beta_2 (k^2)^2 =\beta_2 k^4.
~~~(k_{\mu\nu\rho}  ) (k^{\mu\nu\rho}) = \beta_3 (k^3)^2 =\beta_3 k^6......\eqno (9)$$
where the proportionality factors in eq-(9) are  
the rank and dimension-dependent constants, $\beta_2(D,r=2), \beta_3(D,r=3 ) ....$ associated with the {\bf 2}-vector, 
{\bf 3}-vector,.........components of the polymomentum {\bf K}, respectively. $\beta =1$ for the first term in eq-(7), a rank one tensor : vector.  
The coincidence condition implies that upon setting $\Lambda =0$ all the {\bf p}-loop histories $collapse$ to a point history. 
In that case the areas, volumes, ...hypervolumes 
collapse to $zero$ and the wave equation (5) reduces to the ordinary Klein-Gordon equation for a spin zero massive particle.

Factoring out the  $k^2$ factor in (7), using the analog of the 
dispersion relation (8) and taking the square root, after performing the
binomial/Taylor expansion of the square root, 
subject to the condition $\Lambda^2  k^2 << 1$, 
one obtains an $effective$ energy dependent Planck " constant " that
takes into account the Noncommutative nature of the Clifford manifold 
({\bf C}-space )
at Planck scales :

$$\hbar_{eff} (k^2)  =\hbar (1 +{1\over 2.2!} \beta_2 \Lambda^2 k^2 +
{1\over 2.3!} \beta_3 \Lambda^4  k^4
+...................). \eqno (10) $$
where we have included explicitly the $D$ and rank dependent  coefficients
$\beta_1, \beta_2,\beta_3 ...$ that arise in (9) due to the $coincidence$ condition and on dimensional analysis.

Arguments concerning an effective value of Planck's " constant " related 
to higher
derivative theories and the modified uncertainty relations have been 
given by [8]. 
The advantage of this derivation based on the New  Relativity Principle 
is that one
automatically avoids the problems involving the ad hoc introduction of 
higher derivatives in
Physics ( ghosts, ...) .

The uncertainty relations for the coordinates-momenta follow from the
Heisenberg-Weyl algebraic relation familiar in QM :

$$\Delta x \Delta p \ge | <[{\hat x} , {\hat p} ]> |. ~~~
[{\hat x}  , {\hat p} ] =i\hbar  \eqno (11)$$
Now we have that in {\bf C}-spaces, $x,p$ must not, and should not, 
be interpreted as ordinary vectors of spacetime but as one of the many
$components$ of the Clifford-algebra valued multivectors 
that " coordinatize "  the Noncommutative Clifford Manifold, {\bf 
C}-space.
The Noncommutativity is $encoded$ in the $effective$ value of the 
Planck's " constant "
which $modifies$  the Heisenberg-Weyl $x,p$ algebraic commutation 
relations and, consequently,
generates new uncertainty relations :

$$ \Delta x \Delta p \ge | <[ {\hat x} , {\hat p} ] > | = <\hbar_{eff}> =
\hbar (1 +{1\over 2.2!} \beta_2 \Lambda^2 <k^2> +
{1\over 2.3!} \beta_3 \Lambda^4 <k^4> +.......)    \eqno (12)$$

Using the relations :

$$\hbar k =p.~~~ <p^2 > \ge (\Delta p)^2. ~~~<p^4 > \ge (\Delta 
p)^4.....\eqno (13)$$
one arrives at :

$$\Delta x \Delta p \ge  \hbar + { \beta_2 \Lambda^2 \over 4 \hbar} 
(\Delta p)^2 +
{\beta_3 \Lambda ^4 \over 12 \hbar^3} (\Delta p)^4 +....... \eqno (14)$$

Finally, keeping the first two terms in the expansion in the r.h.s of 
eq- (14)
one recovers the ordinary String Uncertainty Relation [5] directly from 
the New Relativity Principle as promised :

$$\Delta x \ge {\hbar \over \Delta p} + {\beta_2 \Lambda^2 \over 4 
\hbar}(\Delta p).\eqno (15)$$
which is just a reflection of the minimum distance condition in Nature 
[3,4,5,6,7,10] 
and an inherent Noncommutative nature of the Clifford manifold  ( {\bf 
C}-space ). Eq-(15) yields a $minimum$ value of $\Delta x$ of the order of the Planck length $\Lambda$ 
that can be verified explicitly simply by minimizing eq-(15).

\bigskip

\centerline{\bf 3. A Simple Argument Why Quantum Spacetime could be  Infinite Dimensional}
\centerline{\bf A Plausible Resolution of the Cosmological Constant Problem}

\bigskip

So far the derivation of the String uncertainty relations from the New Relativity Principle has been straightforward. 
However, we wish to be more radical in our approach. An immediate question comes to mind :

 $$Why~ did~ we~ truncate~ the~ series~ eqs-(5,7)~ to~ a~ finite~ value~ of~ the~ Quantum~ Spacetime~ dimension ? $$  

If the New Relativity principle is true then we must include {\bf all} dimensions for the Quantum Spacetime. 
It is {\bf all} or {\bf nothing} !  
Taking this radical view will generate instead of the finite series of eq-(7) an $infinite$ series of the form :

$$\hbar^2 k^2 \sum_{r=1}^{\infty} {\beta_r (r,D) \over r!}    (k\Lambda)^{2(r-1)} = lim_{p\rightarrow \infty} {\Lambda^{2p}({\cal E}^2-\lambda^2)  \over \hbar ^{2p}}.  \eqno (16) $$
where $r=1,2,3,......D$ denotes the rank of the vector, {\bf 2}-vector, {\bf 3}-vector,.....associated with the Clifford-algebra valued polymomentum {\bf K} conjugate 
to the Clifford-valued " line " in {\bf C}-space : {\bf X}$(\Sigma)$. 

The sum of the infinite series depends on the infinite family of rank and dimension dependent coefficients $\beta_r (D, r) $ appearing in eq-(9,10,12). 
For simplicity purposes, and for the sake of the argument, we will take $all$ of the coefficients to have the simplest value of them all : {\bf 1}. 
The infinite series yields  :

$$ \sum_{r=1}^{\infty} { (k\Lambda)^{2(r-1)}\over \ r!} =\sum_{r=1}^{\infty} { z^{2(r-1) } \over r!}= 
\sum_{r'=0}^{\infty} { z^{2 r' } \over (r'+1)!} = {e^{z^2} -1 \over z^2}. ~where~ z\equiv k\Lambda. \eqno (17)$$
One recovers one of the confluent hypergeometric functions as the value of the sum. Therefore, after recasting the sum in terms of hyperbolic functions

$$ \hbar^2_{eff} =\hbar^2 {e^{z^2} -1 \over z^2 } \Rightarrow 
\hbar^2_{eff}=\hbar^2 e^{z^2/2} {  sinh (z^2/2)  \over ( z^2/2 ) }.\eqno (18) $$
following the exact same steps as in the previous section one gets the full blown Uncertainty Relations 
for Quantum Spacetime due to 
the contributions of {\bf all} extended objects : $p=0,1,2,.....\infty$ : :

$$\Delta z\equiv \Lambda \Delta k   
\Rightarrow \Delta x \ge  {\sqrt 2} \Lambda ~{ e^{ (\Delta z )^2 /4 } \over (\Delta z )^2 } 
 \sqrt { sinh [ {  (\Delta z)^2 \over 2 } ]  }.  \eqno (19)$$ 

One can verify that the function :

$$x=x(z^2)\equiv   {\sqrt 2}     \Lambda {  e^{ z^2/4 } \over z^2 } 
 \sqrt { sinh ~({ z^2 \over 2} ) }.   \eqno (20)$$ 
after differentiating it and equating it to zero, has a minimum/maximum   for those values of $z_o$ such that satisfy : 

$$tanh [z^2/2] = {z^2 \over 4-z^2}. \eqno (21)$$
When $z=0,\infty\Rightarrow x=\infty$ as expected in eq-(20), when the momentum is $k=0,\infty$. The {\bf minimum} value of $x$ occurs when 

$$ 1.2621 <z_o < 1.2626 ~~~ x_{min}\sim 1.2426~\Lambda. \eqno (22)$$ 
Therefore, for momentum values  precisely of the order of the Planck's momentum : $k_o\sim (z_o/\Lambda)$ that gives $1.262 ~k_P$ ,   one reaches the {\bf minimum} distance
of $1.2426~\Lambda$ ! as it is required from the New Relativity Principle : Polydimensional Covariance and Scale Relativity [2,3].

Of course, one can always tune the infinite number of coefficients (16) in an infinite number of ways to reproduce the Planck scale as the minimum scale for arbitrary 
values of the momentum. A smaller subclass of infinite tuning possibilities  appears when the Planck scale minimum occurs 
$precisely$ at Planck values of the momentum. It is remarkable that 
the simplicity arguments of setting {\bf all} the values of the coefficients to {\bf 1} yields the desired results of attaining a minimum Planck scale for 
Planck values of the momentum. We believe this is {\bf not} a numerical coincidence.

An immediate question arises : if there are many ways of selecting and tuning the coefficients $\beta_r$ in (16) to satisfy the requirements of attaining
minimal Planck scale uncertainty at Planck scale momentum, i.e there is an infinite range of possible values for the 
$[x,p]= i \hbar _{eff}$ commutation relations, is there a physical criteria to select a unique value of $\hbar$ ?

We believe that the answer to this question may lie in Hopf algebraic structures at Planck scales [12]. 
Recently there has been a lot of activity pertaining the  Hopf algebraic
structure  underlying  to perturbative QFT [9] and the numerical " miracles" of the Renormalization Group process.  
As Kreimer has pointed out, the iterated removal of nested divergences while maintining locality has to fulfill combinatorial properties 
summarized by Zimmermann's forest formula. There is an underlying mathematical structure that is in no way accidental. For relations to low dimensional topology, number theory,...we refer to Kreimer et al [9]. The authors [12] have suggested that there is a Planck scale Hopf algebra as a particular example 
of a Noncommutative differential geometry at Planck scales where the 
Planck scale acts as a natural ultraviolet regulator . 
In fact, Majid found a $[x,p]=i\hbar_{eff} $ commutation relation that bears a striking resemblance to 
ours. Thiemann has also argued that Planck scale should serve as natural regulator for matter QFT [10]. Nottale [3] has argued in numerous occasions 
that there is a deep link between the Renormalization Group process, that is based in scaling arguments, and Scale Relativity. 
In fact he has given a $resolution$  dependent effective Planck's constant that Granik recently has shown to agree with eq-(10) up to the first terms [17]. .

The confluent hypergeometric function that results after summing the infinite
series (16) is {\bf no} numerical accident. Gamma functions have long been known to be essential in the dimensional regularization procedures and in Veneziano's formula
that spawned String Theory.  
In addition, the $crux$ of including {\bf all} dimensions in our calculation for the $[x,p]$ commutation relations is 
that one does {\bf not} have to 
truncate/amputate the  $[x,p]$ commutators to a $finite$ number of terms like it was done in  [7]. We have given in (19) the full blown 
Quantum Spacetime Uncertainty relations that are more general than the usual String Uncertainty relations; We are including the effects of {\bf all} 
extended objects !

The main lesson from this numerical exercise is that Quantum Spacetime could be {\bf infinite}  dimensional if we invoke  
the New Relativity principle 
to the fullest potential 
within the context of Noncommutative Clifford manifolds, {\bf C}-spaces and Quantum Groups ( Hopf algebras).  This 
result  that the Quantum Spacetime is infinite-dimensional has been advocated many times by [3,4]  
within the context of Fractals, Scale Relativity  and a Cantorian-Fractal spacetime : a transfinite infinite nested hierarchy  of fractal Cantorian sets of 
infinite dimensionality. Quantum sets have been proposed long ago by Finkelstein 
in the formulation of Quantum Relativity [16]. This straightforward numerical analysis is a strong indication that 
Quantum Spacetime could be infinite-dimensional and that it may indeed be fractal at its very core. Nature {\bf is} Fractal.  
It is not a big surprise that Quantum Spacetime could be as well. 
Being fractal supports the view of 
Majid [12] that Quantum Geometry is a Braided Categorical one. Since a Fractal Quantum Spacetime has fractal dimensions, it follows naturally that it 
should allow for fractional spins, charges, statistics,.. i.e The Quantum Geometric world has Braided Statistics.  

Ordinary real numbers are no longer  useful to describe the infinite dimensional Fractal Quantum Spacetime we are proposing. 
It is meaningless to assume that we 
can meausure a real number to infinite nonperiodic decimal places. It has been speculated for quite some time that due to the minimal Planck length, the 
geometry at Planck scales is Non-Archimedean. Therefore $p$-adic numbers are the natural ones to use at this scale. For a review of the mathematical applications of 
$p$-adic numbers in Physics and Fractals see [13]. For the role of $p$-adics in the construction of TGD see [15].

If Fractal Quantum Spacetime is indeed $infinite$  dimensional we would have to drastically modify our naive perceptions that spacetime has a {\bf fixed} 
dimension [4] and reconsider the validity of the compactification arguments studied so far from higher to low dimensions. Dimensions are $resolution$ dependent 
[3,4]. Instead
we may be obliged to view $D=4$ only as an overall {\bf average} dimension in the same way that the speed of the molecules inside a box at fixed temperature 
is distributed over 
a wide range of velocities and has an average one related to the temperature. El Naschie [4] using a Gamma distribution for the ensemble of dimensions,  
fractal arguments and Astrophysical data results has obtained average dimensions close to $D=4$. 
The same ideas apply  to the observed spacetime signature and to the resolution of the cosmological constant problem. The New Relativity principle treats all 
dimensions and signatures on the same footing.

To finalize we show why the cosmological "constant"  should not be treated as such : it is in the eye of the beholder . 
The key to a plausible and remarakaby simple solution to the cosmological "constant" problem lies within eq-(16) which is nothing
but an generalization of Einstein's relation : $E^2-p^2=m^2$. One simply shifts the cosmological constant term of (16) to the 
left hand side. Upon shifting  it to the "left"  we have :

$$  ({\Lambda \over \hbar})^{2p} \lambda^2  + \hbar^2 k^2 \sum_{r=1}^{\infty} {\beta_r (r,D) \over r!} (k\Lambda)^{2(r-1)} = 
lim_{p\rightarrow \infty} ~ {\Lambda^{2p} {\cal E}^2  \over \hbar ^{2p}}.  \eqno (23) $$
The New Relativity principle which is based on the principles of Polydimensional Covariance [2] which reshufle a string history for a $5$-brane history; 
a $9$-brane history for a $5$-brane history and so forth i.e the New Relativity principle is nothing but taking Chew's bootstrap idea to the heart : 
each $p$-brane is made of {\bf all} the others ! It is in this fashion why {\bf all} dimensions ( and signatures) must be treated on the {\bf same} 
footing. As Nottale and El Naschie have  argued, dimensions are not absolute concepts in Quantum Spacetime, they are resolution dependent. 

The New Relativity Principle ( Polydimensional Covariance [2] )  states  that the r.h.s (23) is truly an invariant ( like the proper time or proper rest mass of a particle) while 
the terms on the l.h.s are just the analogs of the squared-norm of a four-vector $E^2-p^2=m^2$. Therefore, based on this simple analogy we propose that 
$\lambda$ is {\bf not} a constant but instead is just one of the many observer dependent components of the polymomentum multivector {\bf K} referred earlier in section {\bf 2}. 
Hence we have that the combination :

$$({\Lambda \over \hbar})^{2p} \lambda^2  + \hbar^2 k^2 \sum_{r=1}^{\infty} {\beta_r (r,D) \over r!} (k\Lambda)^{2(r-1)} = 
({\Lambda \over \hbar})^{2p} \lambda'^2  + \hbar^2 k'^2 \sum_{r=1}^{\infty} {\beta_r (r,D) \over r!} (k'\Lambda)^{2(r-1)} = ....=
{\Lambda^{2p}  {\cal E}^2  \over \hbar ^{2p}}.  \eqno (24) $$ 
is an invariant of this New Relativity Theory like 

$$E^2- p^2= E'^2 -p'^2 = E''^2 -p''^2 =.......= m^2. \eqno (25)$$
was in Special  Relativity, where the squared of the {\bf maximal} $p$-brane tension,  associated with the spacetime filling $p$-brane, ${\cal E}^2 $ plays identical role to the one 
played by $m^2$ in Einstein's Relativity. Eq-(24) is remarkably {\bf simple}. It relates the {\bf microscopic} world quantities : 
Planck scale, cosmological " constant " on the left, with the {\bf total } Quantum Spacetime Enery per Unit $p$-volume ( Elasticity of Spacetime quoting Zaharov ) 
associated with the Quantum Spacetime-filling maximal $p$-brane ( $p+1 =D$ ) on the right. The {\bf essential}  terms required to {\bf match}  
the left with the right are precisely 
provided by the {\bf infinite}  number of modes associated with the point-history, loop-history, {\bf 2}-loop history, {\bf 3}-loop history,......
{\bf p}-loop history excitations  {\bf OF} the Quantum Spacetime given precisely by    

$$\hbar^2 k^2 \sum_{r=1}^{\infty} {\beta_r (r,D) \over r!} (k\Lambda)^{2(r-1)}. \eqno (26) $$ 
that led to the full blown Quantum Spacetime Uncertainty Relation (19). For interesting work on the cosmological constant we refer to [19,20]. 
What remains is to find out what is the right Hopf Planck scale algebra that 
selects a unique value of the effective Planck "constant ". Perhaps there are several ?

Since in $D=4$ the Planck legth is given by : 

$$\Lambda = \sqrt { \hbar G \over c^3}. \eqno (27)$$
the immediate question, similar to the one proposed long ago by Dirac,  arises : 
isn' it possible that $\hbar,G, c$ are not "constants" in eq-(27) but they could vary in such 
a way as to leave the value of $\Lambda$ invariant ??? 
In the past years there has been a lot of research activity in the Astrophysics community pondering if the speed of light varies in Cosmology [18] . 
Nottale [3] has given another explanation for the resolution of the cosmological constant problem based on Scale Relativity [3]. His argument is essentially  that 
it is meaningless to compare two values of energy densities at two {\bf different} scales without including Scale-Relativistic effects. He explains why 
the  $10^{50}, 10^{60}$ discrepancy is due to the Scale-Relativistic "Lorentz" dilation factors.

It seems that one may be forced to demolish the old established "idols" ( using Finkelstein' terminoly ) of spacetime, dimension, cosmological constant,....in the same way that Relativity and Quantum Mechanics replaced the Cartesian-Newtonian paradigm. As we end the century, it is time perhaps to embrace a new paradigm in Physics that demolishes
the concept of dimension as an idol.  
Number theory, Topology, Fractals, Cantor sets, $p$-adic analyis, QFT, Quantum Groups, Hopf algebras, Noncommutative Geometry......
seem all to be converging in disguised forms at the Planck scale  upon 
looking at Quantum Spacetime with the magnifying glass
of the New Relativity Theory based on Noncommuative {\bf C}-spaces [1] , the principle of Polydimensional Covariance [2] and Scale Relativity [3] 
: a magnifying glass lying deep inside the fuzzy  crystal ball of imagination signaling what it may turn out to be a new paradigm in Physics.

\smallskip

\centerline{\bf Acknowledgements}
\smallskip 

We are indebted to E. Spallucci for a very constructive critical mail correspondence. 
To G. Chapline. A. Granik,  L.Nottale , W. Pezzaglia, M. El Naschie and D. Finkelstein for 
discussions. Finally many thanks to C. Handy for his assistance and encouragement . 
\smallskip  

\centerline{\bf References}

1. C. Castro : " The String Uncertainty Relations follow from the New Relativity Principle " 

hep-th/0001023. " Hints of a New Relativity Principle from $p$-brane 

Quantum Mechanics " hep-th/9912113.

"Towards the Search for the Orgins of $M$ Ttheory........hep-th/9809102.

2. W. Pezzaglia : " Dimensionally Democratic Calculus and Principles of 
Polydimensional

Physics " gr-qc/9912025.

3. L. Nottale : Fractal Spacetime and Microphysics, Towards the Theory 
of Scale Relativity

World Scientific 1992.

L. Nottale : La Relativite dans Tous ses Etats. Hachette Literature. 
Paris. 1999.

4. M. El Naschie : Jour. Chaos, Solitons and Fractals {\bf vol 10} nos. 
2-3 (1999) 567.

5. D. Amati, M. Ciafaloni, G. Veneziano : Phys. Letts {\bf B 197} (1987) 
81.

D. Gross, P. Mende : Phys. Letts {\bf B 197} (1987) 129.

6. L. Garay : Int. Jour. Mod. Phys. {\bf A 10} (1995) 145.

7. A. Kempf, G. Mangano : " Minimal Length Uncertainty and Ultraviolet

Regularization " hep-th/9612084.

G. Amelino-Camelia, J. Lukierski, A. Nowicki : " $\kappa$ deformed 
covariant phase

space and Quantum Gravity Uncertainty Relations " hep-th/9706031.

8. R. Adler, D. Santiago : " On a generalization of Quantum Theory :

Is the Planck Constant Really Constant ?  " hep-th/9908073

9. D. Kreimer, R. Delbourgo : " Using the Hopf Algebra structure of QFT in calculations " 

 : hep-th/9903249. 

A. Connes, D. Kreimer : " Lessons from QFT-Hopf Algebras and Spacetime Geometries " 

hep-th/9904044  

10. T. Thiemann : " Quantum Gravity as the Natural Regulator of Matter Quantum Field Theories :

gr-qc/9705019.

11. A. Connes : Noncommutative Geometry. Academic Press. New York. 1994.

12. S. Majid , R. Oeckl : " Twisting of Quantum Differentials and the Planck Scale Hopf Algebra "

math.QA/9811054.

S. Majid : Foundations of Quantum Group Theory. Cambridge University

Press. 1995. Int. Jour. Mod. Phys {\bf A 5} (1990) 4689.

L.C. Biedenharn, M. A. Lohe    :  Quantum Groups and q-Tensor Algebras . 
World

Scientific. Singapore . 1995.

13. V.S Vladimorov, I. Volovich and E. Zelenov  : $p-Adics~ in~ Mathematical~ Physics$ . World Scientific 1992.   

14. S. Adler : Quaternionic Quantum Mechanics and Quantum Fields .

Oxford, New York. 1995.

15. M. Pitkanen : " $p$-Adic Topological Geometry Dynamics : Mathematical Ideas " 

hep-th/9506097.  

16. D. Finkelstein : $ Quantum ~Relativity~,  A ~synthesis~ of~ the~ ideas~ of~ Einstein~ and~ Heisenberg $ 

Spinger-Verlag 1995.   

17- A. Granik : Private communication. 

18. S. Alexander : " On the Varying Speed of Light in a Brane-Induced FRW Universe " hep-th/9912037.  

19. G. Chapline : " The Vacuum Energy in a Condensate Model of Spacetime " hep-th/9812129.

20. E. Verlinde, H. Verlinde : " On RG Flow, Gravity and the Cosmological Constant " hep-th/9912018.  

\bye